# High Temperature Superconductivity in the Candidate Phases of Solid Hydrogen

Mehmet Dogan[1,2], Sehoon Oh[1,2], Marvin L. Cohen[1,2,*]

[1] Department of Physics, University of California, Berkeley, CA 94720, USA

[2] Materials Sciences Division, Lawrence Berkeley National Laboratory, Berkeley, CA 94720, USA

[*] To whom correspondence should be addressed: mlcohen@berkeley.edu

**Abstract**

As the simplest element in nature, unraveling the phase diagram of hydrogen is a primary task for condensed matter physics. As conjectured many decades ago, in the low-temperature and high-pressure part of the phase diagram, solid hydrogen is expected to become metallic with a high superconducting transition temperature. The metallization may occur *via* band gap closure in the molecular solid or *via* a transition to the atomic solid. Recently, a few experimental studies pushed the achievable pressures into the 400–500 GPa range. There are strong indications that at some pressure in this range metallization via either of these mechanisms occurs, although there are disagreements between experimental reports. Furthermore, there are multiple good candidate crystal phases that have emerged from recent computational and experimental studies which may be realized in upcoming experiments. Therefore, it is crucial to determine the superconducting properties of these candidate phases. In a recent study, we reported the superconducting properties of the *C2/c*-24 phase, which we believe to be a strong candidate for metallization *via* band gap closure [M. Dogan *et al.*, arXiv:2107.03889 (2021)]. Here, we report the superconducting properties of the Cmca-12, Cmca-4 and I4$_1$/amd-2 phases including the anharmonic effects using a Wannier function-based dense *k*-point and *q*-point sampling. We find that the *Cmca*-12 phase has a superconducting transition temperature that rises from 86 K at 400 GPa to 212 K at 500 GPa, whereas the *Cmca*-4 and *I4$_1$/amd*-2 phases show a less pressure-dependent behavior with their $T_c$ in the 74 – 94 K and 307 – 343 K ranges, respectively. These properties can be used to distinguish between crystal phases in future experiments. Understanding superconductivity in pure hydrogen is also important in the study of high-$T_c$ hydrides.

At low temperatures, hydrogen becomes a molecular solid, which was predicted to transition to a metallic atomic crystal in 1935 [1]. In the wake of the development of the Bardeen–Cooper–Schrieffer (BCS) theory of superconductivity, it was proposed in 1968 that this solid atomic hydrogen would have a high superconducting transition temperature [2]. However, although H is the simplest atom and H$_2$ is the simplest molecule, the same is not true for the crystal structure of solid hydrogen. It has proven challenging to determine the crystal structures of the molecular and atomic phases both theoretically and experimentally. Also, the pressure at which the transition to the atomic phase occurs is similarly difficult to predict and turns out to be high enough to be approachable only recently after several decades of advances in diamond anvil cell techniques. To complicate matters even further, it is possible that a series of structural phase transitions occur in solid hydrogen as the pressure increases, instead of a single transition from a molecular to an atomic phase.

On the theoretical front, several candidates for the molecular and atomic phases were discovered in the early 2000s [3–6]. In the following two decades, with the usage of increasingly complex computational methods that allow for more accurate enthalpy comparisons as well as input from experiments, the number of candidates was reduced, and the three molecular phases (*C2/c*-24,

*Cmca*-12, *Cmca*-4) and one atomic phase (*I4$_1$/amd*-2) emerged as the most promising candidates (number after the dash denotes the number of atoms in the unit cell) [7–16]. Solid molecular hydrogen starts out as a semiconductor, and with increasing pressure, several things may occur that would result in metallization: the same crystal phase may become semimetallic and then metallic *via* band gap closure, a structural phase to another molecular phase (which is semimetallic or metallic) may occur, or a structural phase to an atomic phase may occur. Because there are multiple crystal phases within a few meV of each other in enthalpy at the relevant pressures, and the choices about the treatment of the quantum nature of the light hydrogen nuclei can make a difference in these values, computational studies cannot make a definitive prediction as to which of these scenarios should occur. However, computational studies have been able to pin down the several phases that are competitive in the 300 – 500 GPa range.

On the experimental front, the challenge has been increasing the pressure while maintaining the quality of the sample and making measurements. In the 2000s and 2010s, the accessible pressure range gradually reached 400 GPa, which resulted in the observation of black hydrogen (no transmission in the visible range, indicating that the direct band gap is below ~1.5 eV) around 310 – 320 GPa [17–20], and later in the observation of increased conductivity around 350 – 360 GPa, indicating semimetallic behavior [21,22]. Most recently, an experiment by Loubeyre *et al.* [23] provided the most relevant data to date to help us determine the crystal structure in the 150 – 425 GPa range. Through infrared (IR) absorption measurements which track the vibron frequency and the direct electronic band gap, it was reported that the IR-active vibron frequency linearly decreases with pressure from 150 GPa to 425 GPa, indicating that in this pressure range a single phase remains stable. It was also reported that the direct band gap gradually decreases between 360 – 420 GPa, but abruptly drops below the minimum experimentally observable value of ~0.1 eV. There are disagreements between the few experimental groups that conduct high-pressure experiments on solid hydrogen. However, there are also some agreements, specifically about the increased absorption in the 425 – 440 GPa range, *i.e.* the IR measurements around 425 GPa by Dias *et al.* [21] and the Raman measurements around 440 GPa by Eremets *et al.* [22]. Not much has been claimed about what happens beyond 440 GPa, but it is possible that a phase transition to atomic metallic hydrogen occurs around 500 GPa [24,25].

In a previous work [26], we presented a study of the evolution of the IR-active vibron frequencies in the *C2/c*-24 phase calculated in the anharmonic regime, which closely agree with the observations of Loubeyre *et al.* [23] up to 425 GPa. We also showed that the observed changes in the direct band gap can be explained as a series of changes in the band structure as the pressure increases without postulating a structural phase transition at 425 GPa. Thus, hydrogen may remain in the *C2/c*-24 phase beyond 425 GPa, possibly up to 500 GPa. However, it is also possible that as predicted by Dias *et al.* [21], a structural phase transition occurs to a phase with two lower vibron frequencies with a difference of 300 cm$^{-1}$. Both the *Cmca*-12 and *Cmca*-4 phases have IR-active vibrons with approximately matching frequency differences [26]. Finally, the $I4_1/amd$-2 atomic phase is the strongest candidate for the atomic phase and may be the observed crystal structure around 500 GPa [15,24,25]. Among the four phases we have mentioned, the superconducting properties of the phases with the smaller unit cells (*Cmca*-4 and *I4$_1$/amd*-2) have been computationally investigated by others [6,27,28]. Because of the need for very dense *k*- and *q*-point samplings to obtain converged calculations (among other difficulties), the superconducting properties of the larger phases have remained elusive until recently. However, motivated by the developments in hydrogen-rich materials at high pressures that are breaking the records for superconducting temperatures [29–33], we recently reported an investigation into the superconducting properties of the *C2/c*-24 phase [34], and in this study, we focus on the *Cmca*-12 phase, as well as the *Cmca*-4 and *I4$_1$/amd*-2 phases.

Here, we investigate the electronic structure, vibrational properties, electron–phonon coupling and superconducting properties of solid hydrogen in the *Cmca*-12, *Cmca*-4 and *I4$_1$/amd*-2 phases at 400, 450 and 500 GPa using density functional theory (DFT) calculations in the generalized gradient approximation [35–41], anharmonic corrections with the self-consistent phonon approach [42,43] and a Wannier function-based dense $k$-point and $q$-point sampling [44] (see the **Supplemental Material** for details).

The atomic structures of all three phases are shown in **Figure S1**. The molecular phases *Cmca*-12 and *Cmca*-4 consist of van der Waals-bonded layers of $H_2$ molecules. The atomic phase consists of a highly symmetric 2-atom cell. The electronic structures of the three phases differ qualitatively. Their Fermi surfaces and band structures at 500 GPa are shown in **Figure 1**. At lower pressures, the *Cmca*-12 phase starts out is a semimetal, and the Fermi surface gets larger with new added sheets as the pressure increases [26]. The density of states at the Fermi energy is 0.013, 0.018 and 0.024 states/eV/atom for 400, 450 and 500 GPa respectively. By contrast, the *Cmca*-4 and *I4$_1$/amd*-2 phases have bands that cover both the valence and conduction bands, and their Fermi surfaces remain approximately fixed with increased pressure (~0.020 and ~0.036 states/eV/atom for *Cmca*-4 and *I4$_1$/amd*-2, respectively). We note here that the generalized gradient approximation (GGA) underestimates band gaps. A better treatment of excited states such as the GW approximation can increase band gaps by ~1.5 eV in this pressure range [7,10] and should have a significant effect on the *Cmca*-12 phase. Nuclear quantum effects have the opposite effect reducing band gaps by smearing the position of the potential wells, and this effect happens to be at similar magnitude [45,46]. As a result of this accidental cancellation, we expect the computed electronic structures to be a faithful approximation [26].

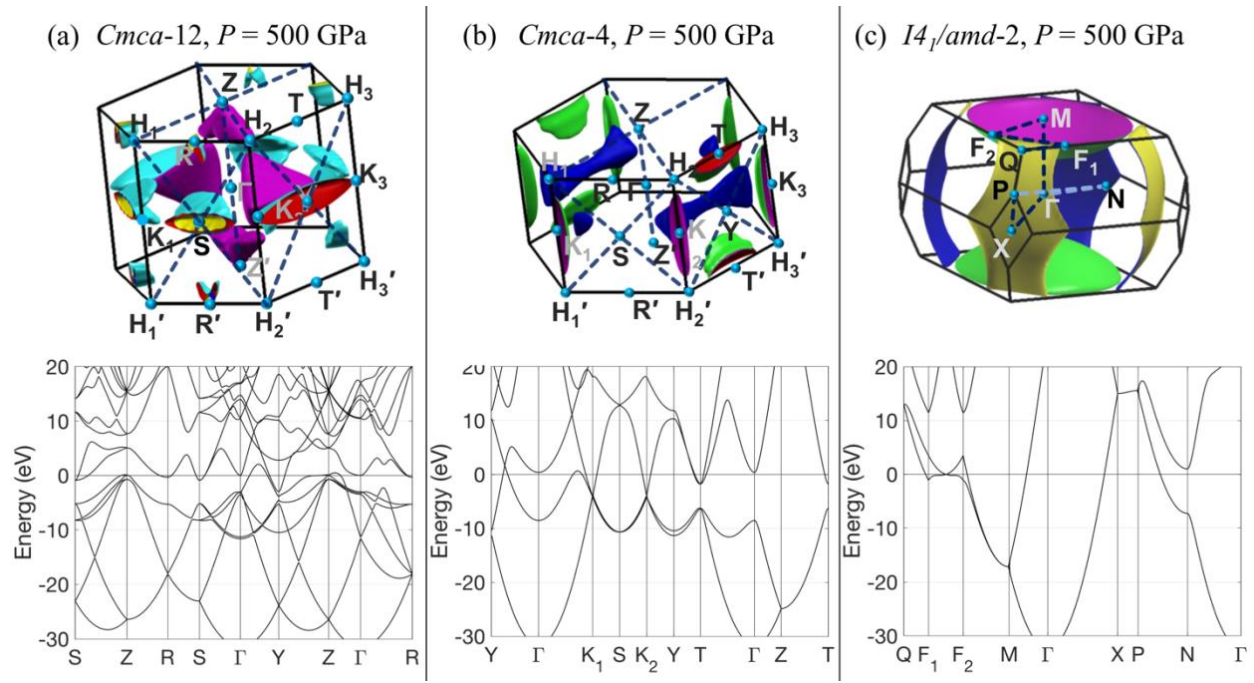

**Figure 1. Fermi surfaces and band structures at 500 GPa. (a)** The first Brillouin zone and the Fermi surface of the *Cmca*-12 phase of hydrogen at 500 GPa, with high-symmetry points labeled, followed by the band structure. Energies are relative to the Fermi energy. **(b)** The same as (a) but for the *Cmca*-4 phase. **(c)** The same as (a) but for the *I4$_1$/amd*-2 phase.

We present the phonon dispersion relations of the *Cmca*-12 phase at 500 GPa both in the harmonic and anharmonic approximations in **Figure 2**. Similar to the *C2/c*-24 phase [34], the anharmonic effects increase the lower phonon frequencies, which correspond to collective motion of the atoms within a plane, and they decrease the higher vibron frequencies. The Eliashberg function ($\alpha^2 F$) and the phonon densities of states (PhDOS) are also presented in **Figure 2**. The fact that $\alpha^2 F$ and PhDOS are qualitatively similar indicates that the electron–phonon coupling is not disproportionately distributed among phonon modes. The electron–phonon coupling parameter defined as $\lambda(\omega) = \int_0^\omega \frac{d\omega'}{\omega'} \alpha^2 F(\omega')$ is also shown in the middle panel of **Figure 2**. We observe that the overall shape of $\lambda(\omega)$ in the harmonic and anharmonic cases are similar, and the fact that the anharmonic values are smaller can be explained by the increase in the lower phonon frequencies, which enter the integral as the denominator. The electron–phonon coupling constant $\lambda$ corresponds to $\lambda(\infty)$ and is equal to 1.49 and 1.35 for the harmonic and anharmonic cases, respectively. The analogous plots for 400 and 450 GPa are presented in **Figure S2** and **Figure S3**, respectively, and show the same general features.

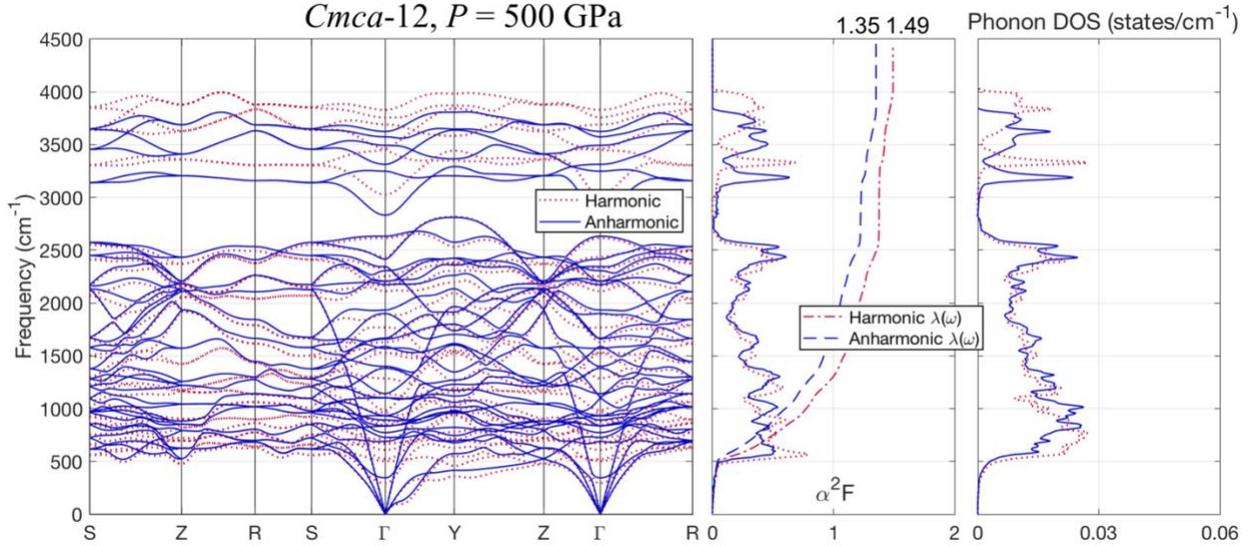

**Figure 2. Phonons and electron–phonon coupling for *Cmca*-12 at 500 GPa.** The phonon dispersion relations for the *Cmca*-12 phase of hydrogen at 500 GPa pressure (left panel). The harmonic and anharmonic calculations are shown by red dashed lines and blue solid lines, respectively. The Eliashberg function $\alpha^2 F$ and the electron–phonon coupling parameter $\lambda(\omega)$ (middle panel), and the phonon densities of states (right panel).

In order to compare it with the *Cmca*-12 phase as well as previous studies, we next investigate the phonon spectra of the *Cmca*-4 phase at 500 GPa both in the harmonic and anharmonic approximations, shown in **Figure 3**. In this phase, the anharmonic effects generally squeeze the overall dispersion of phonons and slightly reduce the electron–phonon coupling. Comparison between $\alpha^2 F$ and PhDOS indicates that the higher frequency phonons, which correspond to intramolecular vibrations, boost the electron–phonon coupling. The analogous plots for 400 and 450 GPa are presented in **Figure S4** and **Figure S5**, respectively, and show the same general features. We notice that for higher pressures, especially for 500 GPa, anharmonic corrections significantly soften some low frequency modes, indicating that the system is becoming dynamically unstable. We note that our computed $\lambda$ values differ from those of Cudazzo *et al.* who used superconducting density functional theory (SCDFT) to find that $\lambda$ doubles from ~1 at 400 GPa to ~2 at 450 GPa [6,47,48]. A more recent study by Borinaga *et al.* [27] found that $\lambda = 0.89$ in the harmonic approximation, which is close to our $\lambda = 0.86$. However, their reported anharmonic $\lambda$ is 2.00, which is much higher than our result, *i.e.* 0.77. This is likely due to the Fermi surface pockets around the $\Gamma$ and R points found in these studies and not in the present study. Although the electronic and vibrational properties at the DFT and DFPT levels are computed to a higher precision in our study, we apply anharmonic corrections directly onto the static nuclear coordinates without attempting to modify these coordinates based on zero-point motion effects. Ultimately it is difficult to know if these Fermi surface pockets occur in the actual material, but if they do, then the superconducting properties should be enhanced as predicted by these earlier studies.

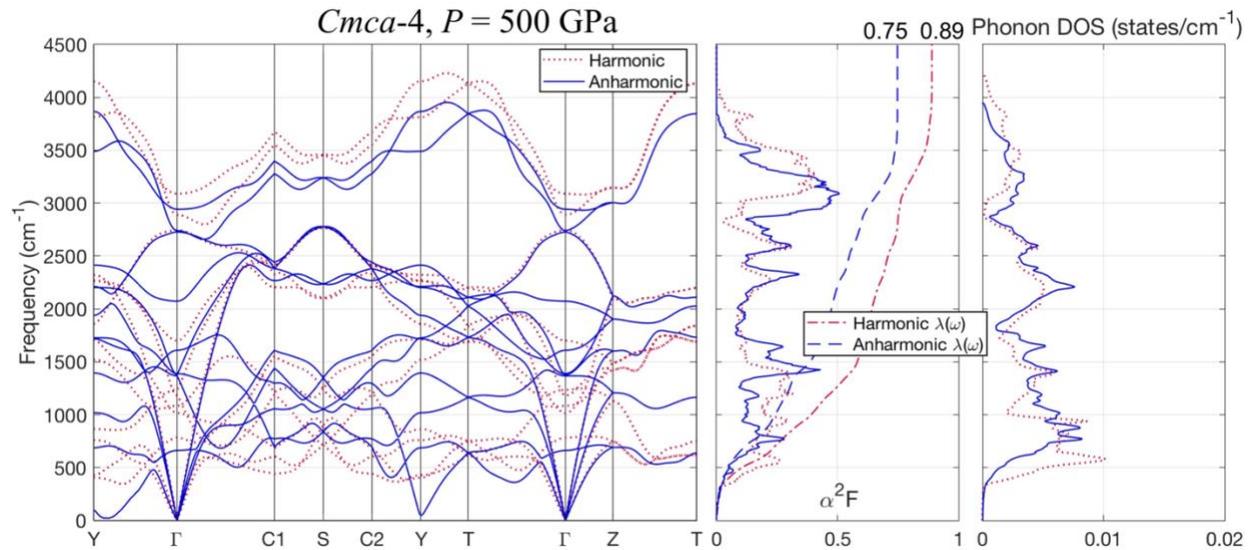

**Figure 3. Phonons and electron–phonon coupling for Cmca-4 at 500 GPa.** The phonon dispersion relations for the Cmca-4 phase of hydrogen at 500 GPa pressure (left panel). The harmonic and anharmonic calculations are shown by red dashed lines and blue solid lines, respectively. The Eliashberg function $\alpha^2 F$ and the electron–phonon coupling parameter $\lambda(\omega)$ (middle panel), and the phonon densities of states (right panel).

The phonon spectra of the *I4₁/amd*-2 phase at 500 GPa both in the harmonic and anharmonic approximations are presented in **Figure 4**. We find that the anharmonic effects generally push up the phonon frequencies and reduce the electron–phonon coupling. The analogous plots for 400 and 450 GPa are presented in **Figure S4** and **Figure S5**, respectively, and show the same general features. Our results for the *I4$_1$/amd*-2 phase generally agree with the previous work by McMahon and Ceperley [49], and also with Borinaga *et al.* [28] who also found suppression of superconducting properties by anharmonicity.

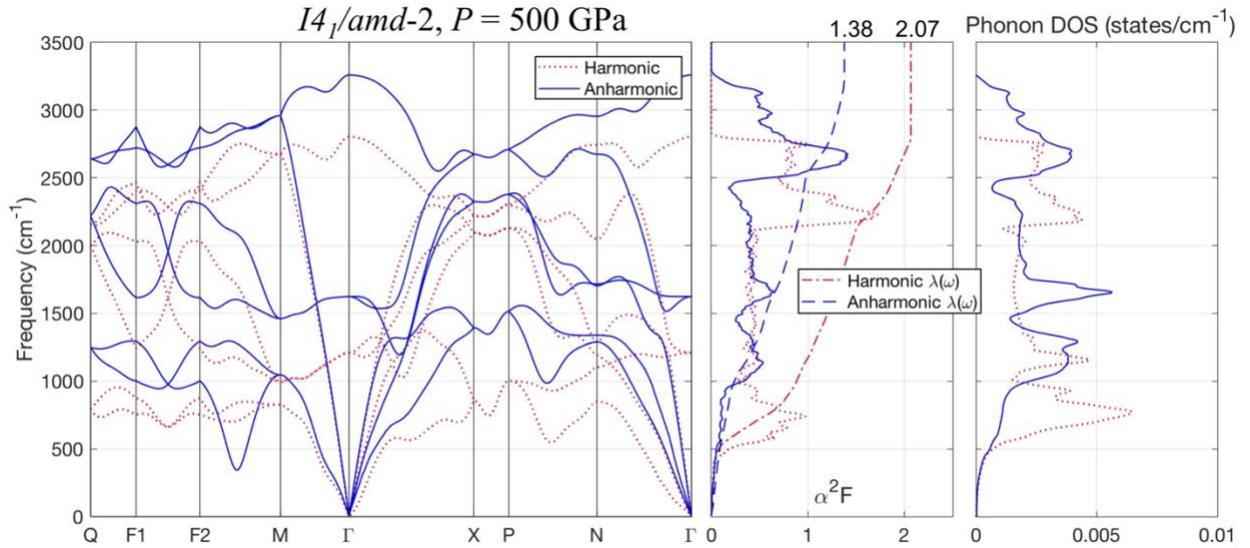

**Figure 4. Phonons and electron–phonon coupling for *I4$_1$/amd*-2 at 500 GPa.** The phonon dispersion relations for the *I4$_1$/amd*-2 phase of hydrogen at 500 GPa pressure (left panel). The harmonic and anharmonic calculations are shown by red dashed lines and blue solid lines, respectively. The Eliashberg function $\alpha^2 F$ and the electron–phonon coupling parameter $\lambda(\omega)$ (middle panel), and the phonon densities of states (right panel).

All of the calculated electron–phonon coupling parameters ($\lambda$) and superconducting transition temperatures ($T_c$) resulting from the Allen–Dynes formula [50] ($\mu^* = 0.1$) are reported in **Table 1**. We see that for the *Cmca*-12 phase, the $T_c$ values for harmonic and anharmonic cases are closer than the differences between harmonic and anharmonic $\lambda$, which is due to the $\omega_{log}$ values having the opposite trend. For the other two phases, the anharmonic effects reduce $\lambda$ and $T_c$ by similar amounts.

**Table 1. Electron–phonon coupling constant and $T_c$.** The electron–phonon coupling constant $\lambda$ and the superconducting transition temperature $T_c$ using both the Allen–Dynes formula and the isotropic Eliashberg theory are shown in the harmonic and anharmonic cases for 400, 450 and 500 GPa for the *Cmca*-12, *Cmca*-4 and *I4$_1$/amd*-2 phases. The Coulomb pseudopotential $\mu^*$ is set to 0.1 in all cases.

| Phase | $P$ (GPa) | $\lambda$ | | $T_c$ (Allen–Dynes) (K) | | $T_c$ (Eliashberg) (K) | |
|---|---|---|---|---|---|---|---|
| | | harmonic | anharmonic | harmonic | anharmonic | harmonic | anharmonic |
| *Cmca*-12 | 400 | 0.87 | 0.79 | 81 | 72 | 96 | 86 |
| | 450 | 1.24 | 1.17 | 142 | 138 | 169 | 161 |
| | 500 | 1.48 | 1.35 | 174 | 172 | 211 | 212 |
| *Cmca*-4 | 400 | 0.76 | 0.67 | 74 | 62 | 78 | 74 |
| | 450 | 0.86 | 0.77 | 94 | 83 | 116 | 94 |
| | 500 | 0.89 | 0.75 | 96 | 81 | 111 | 85 |
| *I4$_1$/amd*-2 | 400 | 2.18 | 1.73 | 262 | 239 | 359 | 329 |
| | 450 | 2.16 | 1.79 | 261 | 220 | 366 | 343 |
| | 500 | 2.07 | 1.38 | 254 | 224 | 348 | 307 |

It is well known that the Allen–Dynes formula underestimates $T_c$ in this class of materials [34,51]. We thus proceed to the isotropic Eliashberg calculations to determine the leading edge of the superconducting gap ($\Delta_0$) *vs.* temperature. Our resulting $T_c$ values are presented in **Table 1** and **Figure 5(a)**, and a representative calculation at 500 GPa is presented in **Figure 5(b).** In **Figure 5(a)**, the Eliashberg $T_c$ values taken from our previous study of the *C2/c*-24 are also presented for completeness. $T_c$ rapidly increases with pressure for the phases where the electronic density of states at the Fermi level also increases with pressure. These values and their pressure dependence can be used in experiments to distinguish between crystal phases.

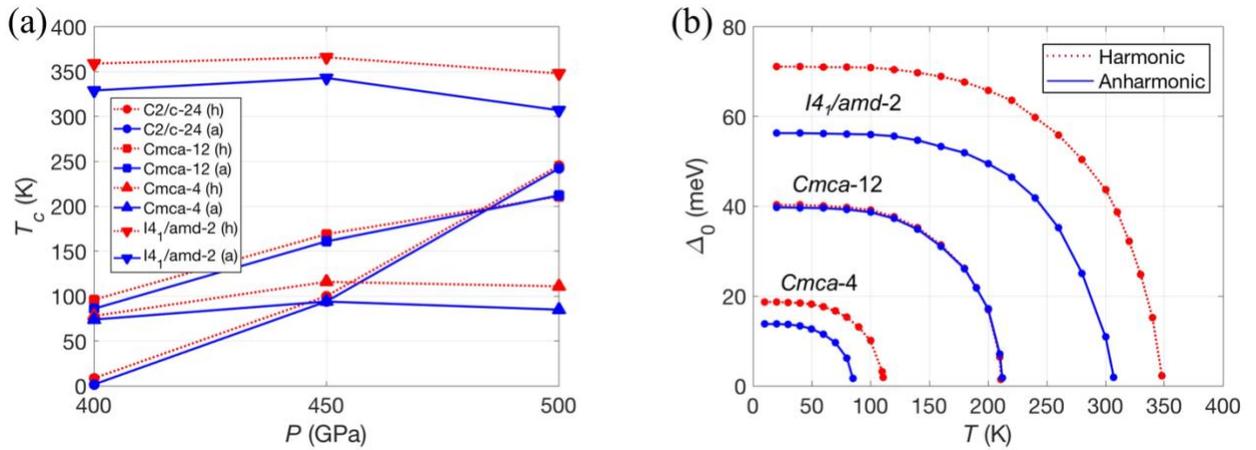

**Figure 5. $T_c$ and superconducting gap. (a)** Superconducting transition temperature computed *via* isotropic Eliashberg equations for the *Cmca*-12, *Cmca*-4 and *I4$_1$/amd*-2 phases and the previously calculated *C2/c*-24 phase [34]. **(b)** Leading edge of the superconducting gap *vs.* temperature for the *Cmca*-12, *Cmca*-4 and *I4$_1$/amd*-2 phases at 500 GPa in the harmonic and anharmonic approximations.

Recently, theoretical efforts have been focused on proposing formulae that can predict superconducting properties from simple calculations. In this vein, a parameter called the networking value ($\phi$) based on the electron localization function (ELF) was proposed as highly correlated with $T_c$ in hydrides [51]. The parameter $\phi$ is obtained by finding the largest value of ELF for which the isosurface of ELF creates a connected network in all three spatial directions. In **Figure S1(a-c)**, we present the plot of the ELF isosurface at the networking value for the three phases in question at 500 GPa. Plotting all the anharmonic Eliashberg $T_c$ and $\phi$ values for the three phases plus the *C2/c*-24 phase [34], we arrive at **Figure S1(d)**. The linear fit results in a weak correlation with an $R^2$ value of 0.58. However, when we plot $T_c$ vs. DOS@$E_{Fermi}$ (**Figure S1(e)**), the correlation significantly improves to an $R^2$ value to 0.89. When many hydrides are compared (177 hydrides in ref [51]), $\phi$ provides a better correlation with $T_c$ than DOS@$E_{Fermi}$, but we conclude that this measure cannot be used in a smaller sample, in particular, to compare the different crystal structures of the same hydride (in this case, hydrogen). As a final note, we present the dielectric function and reflectivity of the *Cmca*-12, *Cmca*-4 and *I4$_1$/amd*-2 phases at 500 GPa calculated in the random phase approximation (RPA) in **Figure S8**. Because it is very challenging to characterize these crystal structures using other methods, these additional data points may be useful for experiments that measure optical properties of high-pressure hydrogen samples.

In summary, we have found that the predicted superconducting transition temperatures for the *Cmca*-12 phase at 400, 450 and 500 GPa are 86, 161 and 212 K, respectively. This rise of $T_c$ with pressure is explained by the increase in the density of states at the Fermi level, which parallels the previously reported *C2/c*-24 phase [34]. On the other hand, $T_c$ has a more modest dependence on pressure for the *Cmca*-4 and *I4$_1$/amd*-2 phases. We have also found that anharmonic corrections cause a slight to moderate reduction in the electron–phonon coupling and $T_c$ for all the studied phases. The current status of the theoretical and experimental research on solid hydrogen requires us to take all of these four crystal phases seriously as candidates in the 300 – 500 GPa range, with special attention given to the *C2/c*-24 and *Cmca*-12 phases. Our results indicate that by measuring superconducting properties on high-pressure hydrogen, experimental researchers can also determine the crystal structure, bringing us one step closer to elucidating the properties of the simplest material in nature under extreme conditions.

**Acknowledgements**

This work was supported primarily by the Director, Office of Science, Office of Basic Energy Sciences, Materials Sciences and Engineering Division, of the U.S. Department of Energy under contract No. DE-AC02-05-CH11231, within the Theory of Materials program (KC2301), which supported the structure optimization and calculation of vibrational properties. Further support was provided by the NSF Grant No. DMR-1926004 which supported the determination of electron–phonon interactions. Computational resources used were Cori at National Energy Research Scientific Computing Center (NERSC), which is supported by the Office of Science of the US Department of Energy under contract no. DE-AC02-05CH11231, Stampede2 at the Texas Advanced Computing Center (TACC) through Extreme Science and Engineering Discovery Environment (XSEDE), which is supported by National Science Foundation (NSF) under grant no. ACI-1053575, Frontera at TACC, which is supported by NSF grant no. OAC-1818253, and Bridges-2 at the Pittsburgh Supercomputing Center (PSC), which is supported by NSF award number ACI-1928147. We thank Hyungjun Lee for technical assistance with the EPW code.

# Supplemental Material for "High Temperature Superconductivity in the Candidate Phases of Solid Hydrogen"

Mehmet Dogan[1,2], Sehoon Oh[1,2,3], Marvin L. Cohen[1,2,*]

[1] Department of Physics, University of California, Berkeley, CA 94720, USA
[2] Materials Sciences Division, Lawrence Berkeley National Laboratory, Berkeley, CA 94720, USA
[*] To whom correspondence should be addressed: mlcohen@berkeley.edu

**Computational Methods**

We compute optimized crystal structures using density functional theory (DFT) in the Perdew–Burke–Ernzerhof generalized gradient approximation (PBE GGA), [1] using the QUANTUM ESPRESSO software package. We use norm-conserving pseudopotentials with a 120 Ry plane-wave energy cutoff. [2,3] The Monkhorst–Pack $k$-point meshes we use to sample the Brillouin zone are listed in **Table S1**. All atomic coordinates are relaxed until the forces on all the atoms are less than $10^{-4}$ Ry/$a_0$ in all three Cartesian directions ($a_0$: Bohr radius). After each structure has been relaxed at a given pressure, a denser sampling of the Brillouin zone is examined to determine the band structure, using a k-point mesh that is twice as fine in all three reciprocal lattice directions. The Fermi surfaces are plotted using the XCrySDen package. [4]

Using density functional perturbation theory (DFPT), [5,6] we calculate the vibrational modes in the harmonic approximation for the pressures of 400, 450 and 500 GPa. Full phonon dispersions are calculated using the $q$-point samplings listed in **Table S1**. We apply the anharmonic corrections using the self-consistent phonon method as implemented in the ALAMODE package. [7,8] For the anharmonic corrections, we use the finite-displacement approach with the real-space supercells listed in **Table S1**. [9] The displacements from the equilibrium position are 0.01, 0.03, and 0.04 Å for the second, third, and fourth order terms of the interatomic force constant, respectively. We use the EPW code to implement the Wannier function-based $k$-space and $q$-space sampling for electron–phonon coupling calculations as well as isotropic Eliashberg theory calculations. [10] The Wannier interpolation is done using fine $k$-point and $q$-point samplings that are given in **Table S1**, using Wannierization with H 1s orbitals as initial projections, (–30 eV, 10 eV) as the inner window and (–30 eV, 70 eV) as the outer window around the Fermi energy. To demonstrate the accuracy of the interpolation in the inner window, the Wannier-interpolated band structures along with the DFT band structure at 400 GPa are presented in **Figure S9**. The comparison is essentially the same for the other pressures.

**Table S1. Sampling parameters for all calculations.** All the sampling parameters for the calculations of the *Cmca*-12, *Cmca*-4 and *I4$_1$/amd*-2 phases.

| Calculation stage | *Cmca*-12 | *Cmca*-4 | *I4$_1$/amd*-2 |
|---|---|---|---|
| *k*-point mesh for DFT | 16×16×16 | 32×32×16 | 32×32×32 |
| *q*-point mesh for DFPT | 4×4×4 | 8×8×4 | 8×8×8 |
| self-consistent phonon supercell | 2×2×2 | 4×4×2 | 4×4×4 |
| fine *k*-point mesh for e–ph coupling | 32×32×32 | 64×64×32 | 64×64×64 |
| fine *q*-point mesh for e–ph coupling | 8×8×8 | 16×16×8 | 16×16×16 |

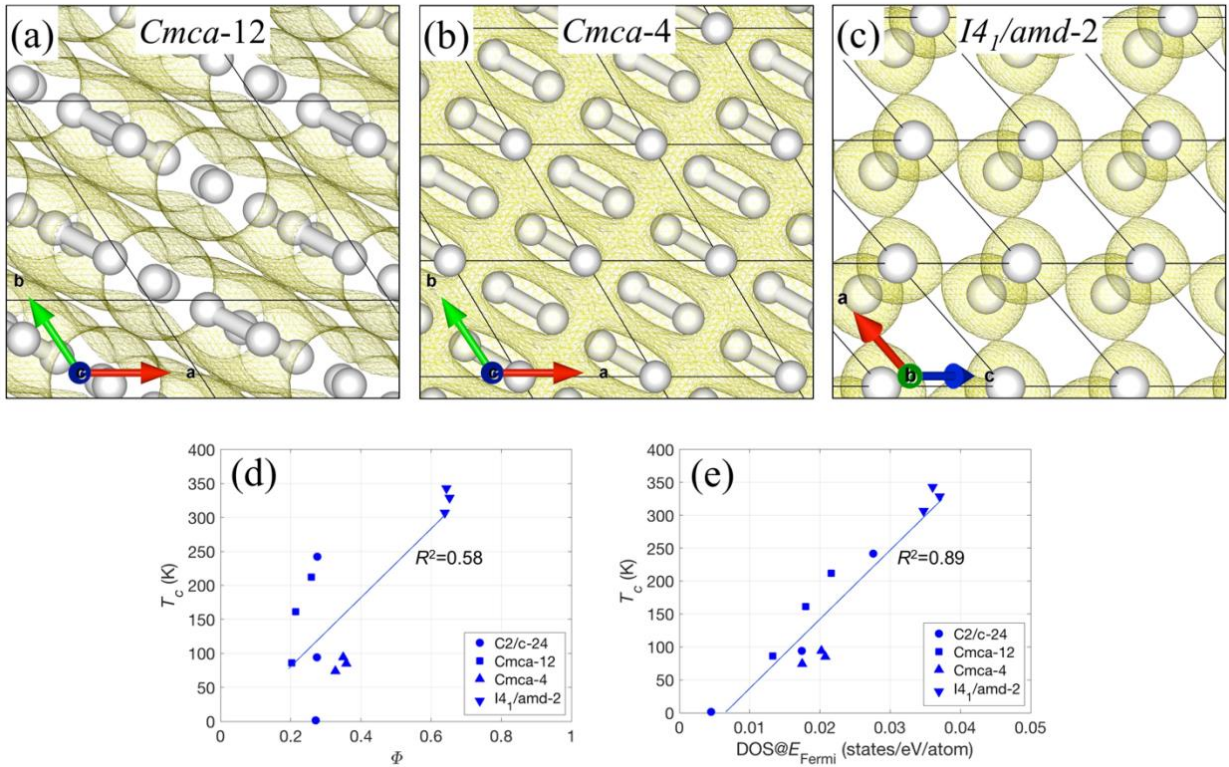

**Figure S1. Atomic structures, ELF isosurfaces taken at the networking values and dependence of $T_c$ on networking value. (a-c)** The atomic structures of the *Cmca-12,* Cmca-*4 and* I4$_1$/amd-*2* phases at 500 GPa, respectively. The unit cell is also shown as a frame. The electron localization function (ELF) is plotted in the figures as isosurfaces with the networking value Φ (see main text for details). **(d)** $T_c$ vs. Φ for the calculated phases as well as the previously reported *C2/c*-24 phase. **(e)** $T_c$ vs. the density of states at the Fermi energy for the calculated phases as well as the previously reported *C2/c*-24 phase.

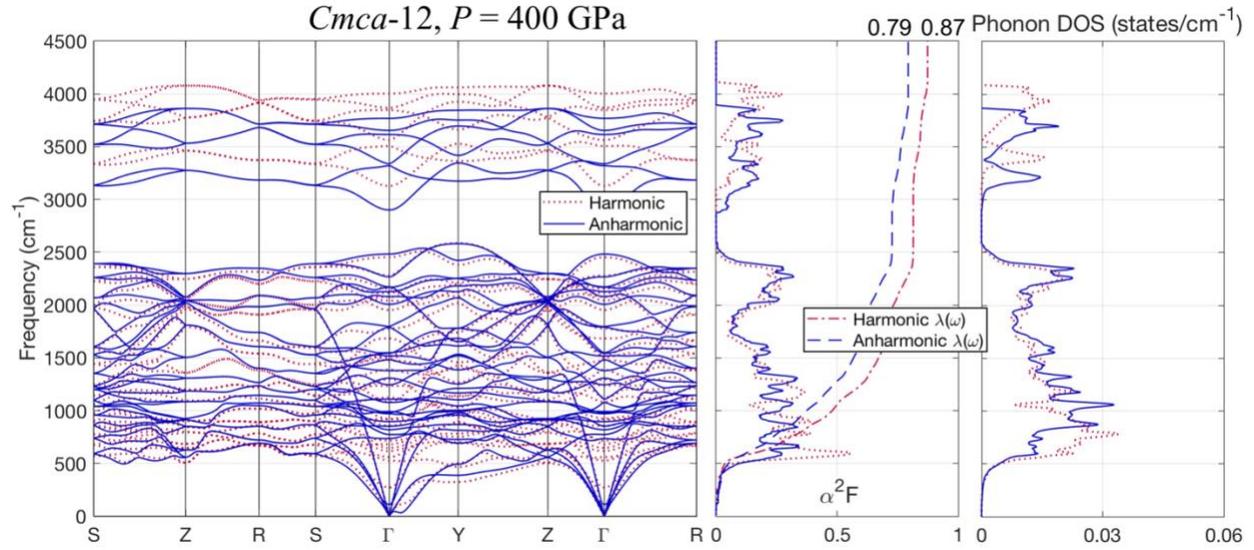

**Figure S2. Phonons and electron–phonon coupling for *Cmca*-12 at 400 GPa.** The phonon dispersion relations for the *Cmca*-12 phase of hydrogen at 400 GPa pressure (left panel). The harmonic and anharmonic calculations are shown by red dashed lines and blue solid lines, respectively. The Eliashberg function $\alpha^2 F$ and the electron–phonon coupling parameter $\lambda(\omega)$ (middle panel), and the phonon densities of states (right panel).

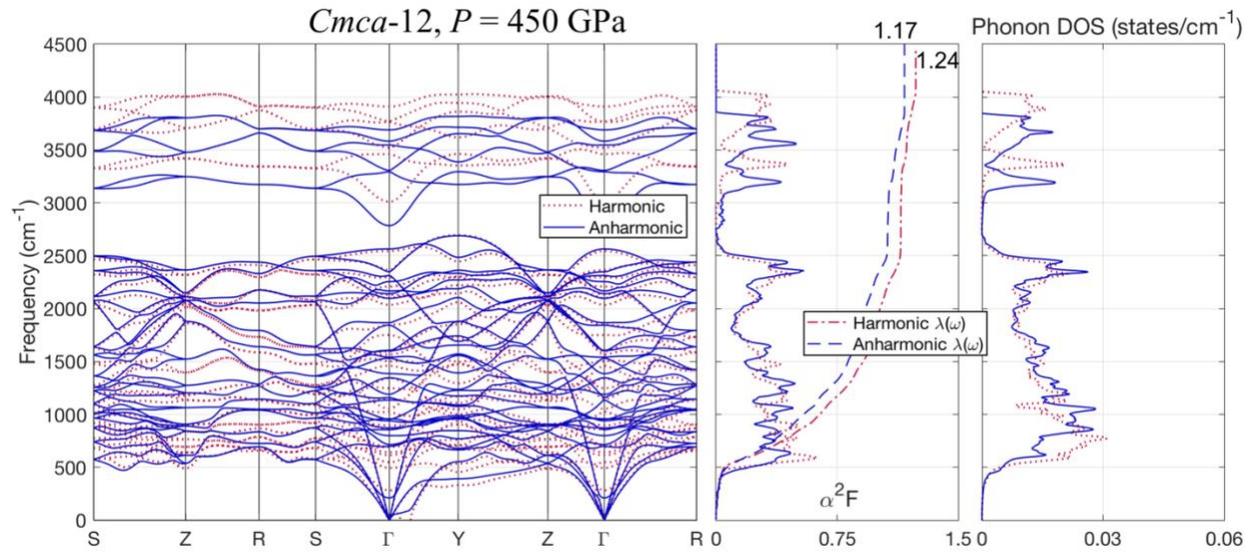

**Figure S3. Phonons and electron–phonon coupling for *Cmca*-12 at 450 GPa.** The phonon dispersion relations for the *Cmca*-12 phase of hydrogen at 450 GPa pressure (left panel). The harmonic and anharmonic calculations are shown by red dashed lines and blue solid lines, respectively. The Eliashberg function $\alpha^2 F$ and the electron–phonon coupling parameter $\lambda(\omega)$ (middle panel), and the phonon densities of states (right panel).

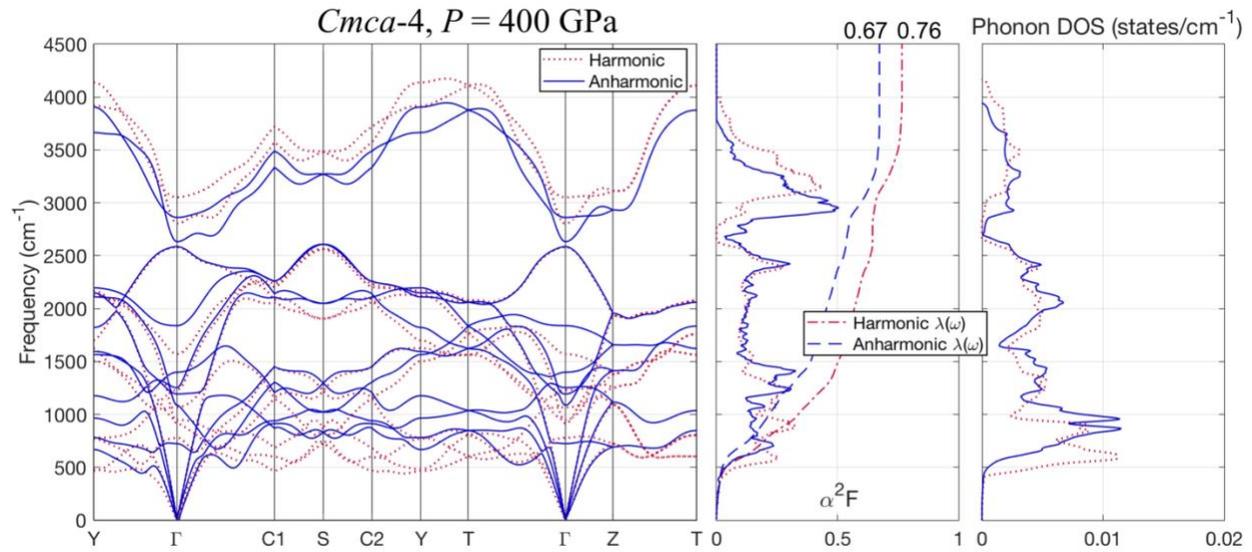

**Figure S4. Phonons and electron–phonon coupling for *Cmca*-4 at 400 GPa.** The phonon dispersion relations for the *Cmca*-4 phase of hydrogen at 400 GPa pressure (left panel). The harmonic and anharmonic calculations are shown by red dashed lines and blue solid lines, respectively. The Eliashberg function $\alpha^2 F$ and the electron–phonon coupling parameter $\lambda(\omega)$ (middle panel), and the phonon densities of states (right panel).

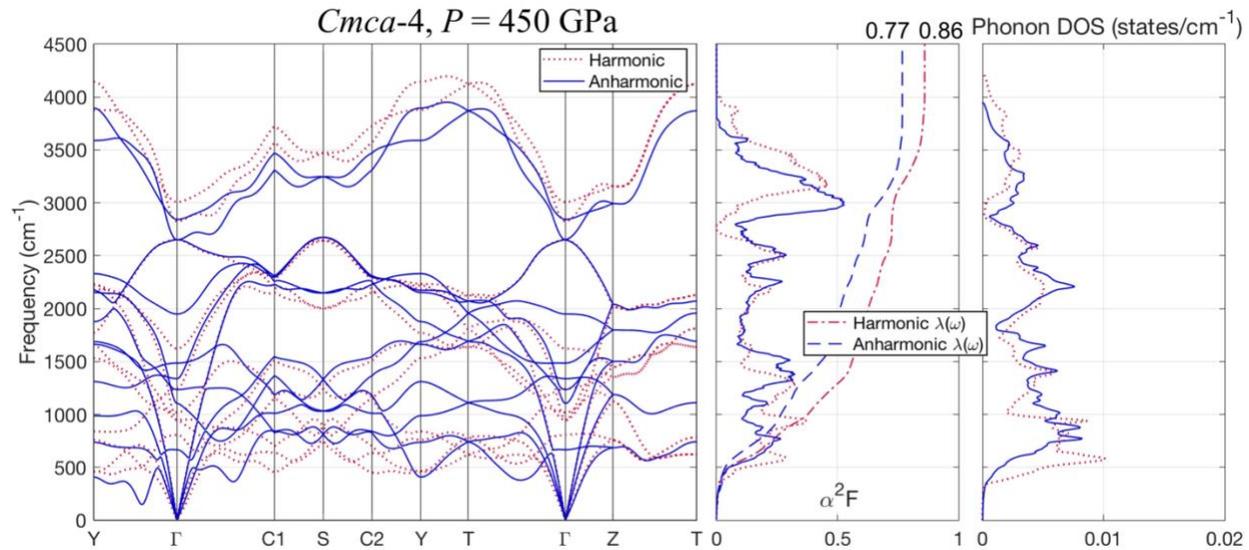

**Figure S5. Phonons and electron–phonon coupling for *Cmca*-4 at 450 GPa.** The phonon dispersion relations for the *Cmca*-4 phase of hydrogen at 450 GPa pressure (left panel). The harmonic and anharmonic calculations are shown by red dashed lines and blue solid lines, respectively. The Eliashberg function $\alpha^2 F$ and the electron–phonon coupling parameter $\lambda(\omega)$ (middle panel), and the phonon densities of states (right panel).

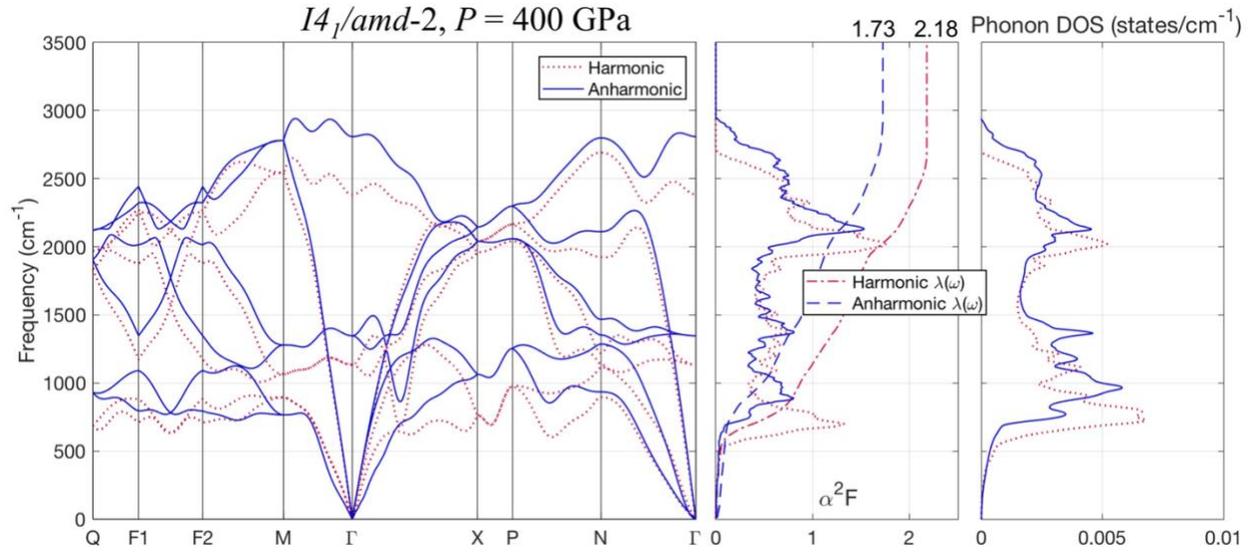

**Figure S6. Phonons and electron–phonon coupling for $I4_1/amd$-2 at 400 GPa.** The phonon dispersion relations for the $I4_1/amd$-2 phase of hydrogen at 400 GPa pressure (left panel). The harmonic and anharmonic calculations are shown by red dashed lines and blue solid lines, respectively. The Eliashberg function $\alpha^2 F$ and the electron–phonon coupling parameter $\lambda(\omega)$ (middle panel), and the phonon densities of states (right panel).

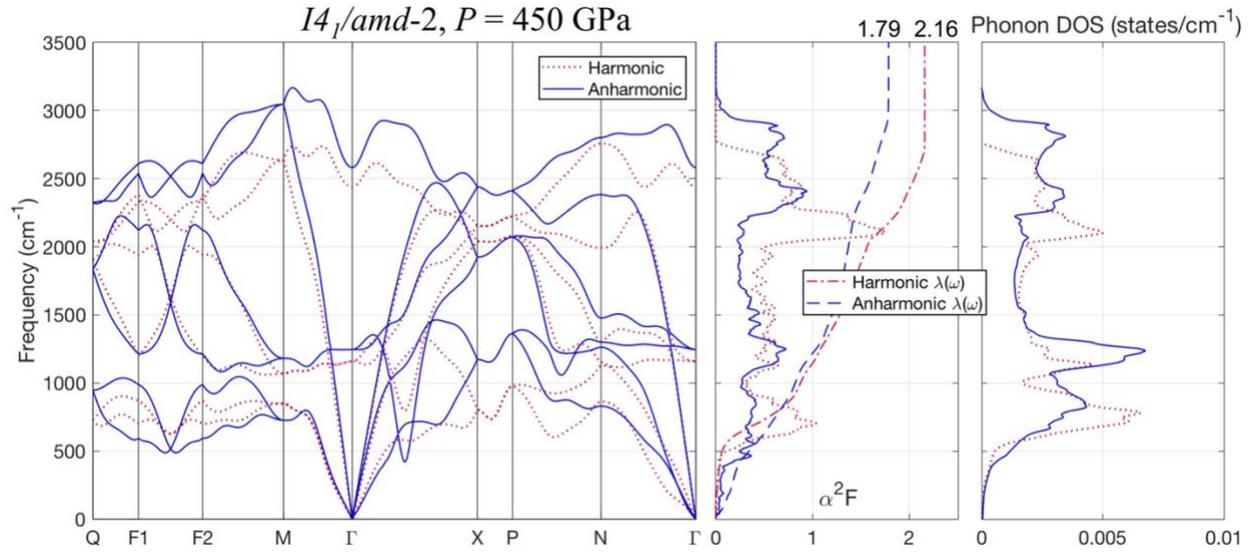

**Figure S7. Phonons and electron–phonon coupling for $I4_1/amd$-2 at 450 GPa.** The phonon dispersion relations for the $I4_1/amd$-2 phase of hydrogen at 450 GPa pressure (left panel). The harmonic and anharmonic calculations are shown by red dashed lines and blue solid lines, respectively. The Eliashberg function $\alpha^2 F$ and the electron–phonon coupling parameter $\lambda(\omega)$ (middle panel), and the phonon densities of states (right panel).

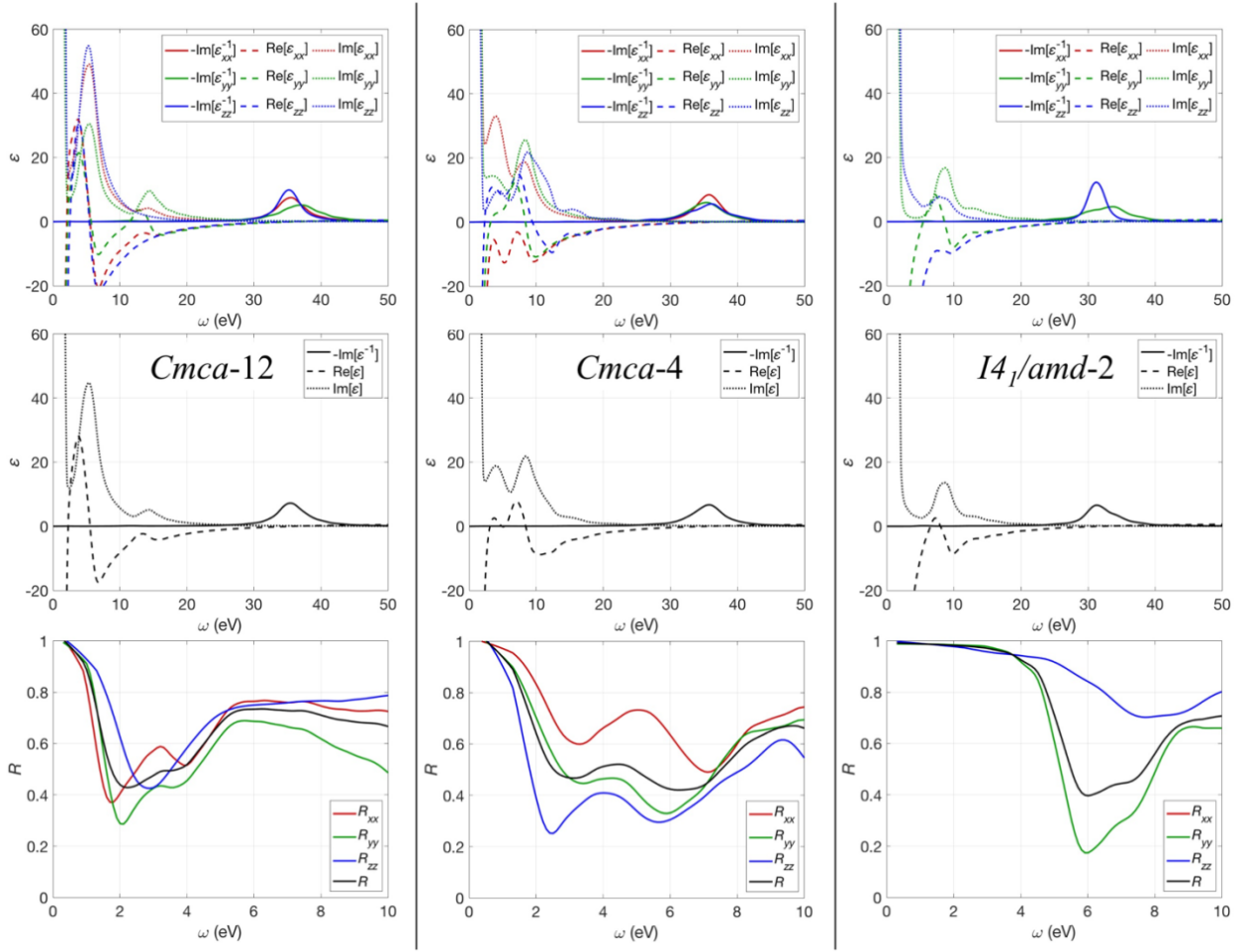

**Figure S8. Dielectric function and reflectivity at 500 GPa.** Dielectric function and reflectivity of the *Cmca*-12 (left), *Cmca*-4 (middle) and *I4$_1$/amd*-2 (right) phases of hydrogen at 500 GPa in the random phase approximation (RPA). In the top row, the components of $\varepsilon$ are plotted separately. In the middle row, the components of $\varepsilon$ are averaged. In the bottom row, reflectivity components are plotted in color and average reflectivity is plotted in black.

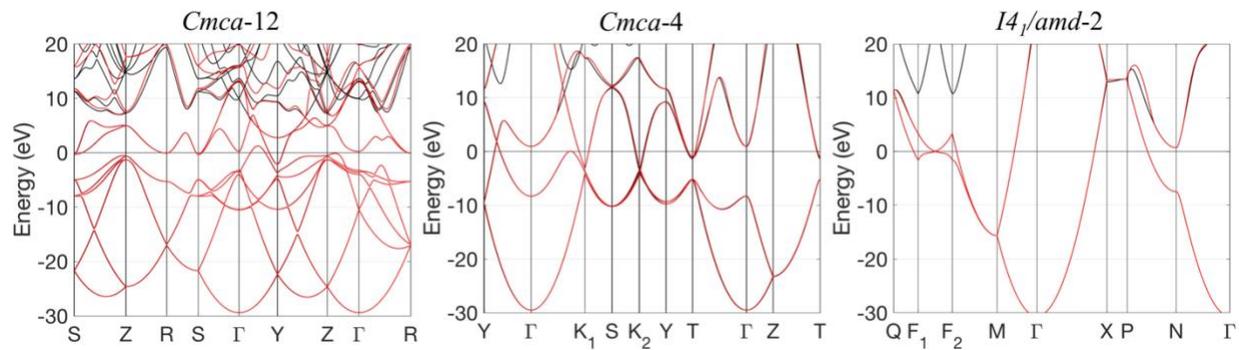

**Figure S9. Original and Wannier-interpolated band structures at 400 GPa.** The Wannier-interpolated band structure (red) overlaid on the original DFT-calculated band structure (black) at 400 GPa for all three phases.